\documentclass[10pt,showpacs,floatfix,letterpaper,amsmath,amsfonts,amssymb,aps,pra,reprint,superscriptaddress]{revtex4-1}
\usepackage[latin1]{inputenc}
\usepackage{bm}
\usepackage{graphicx}
\usepackage{tensor}
\usepackage{epstopdf}

\newcommand{\op}[1]{\hat{#1}}                                 % Predictive Operator
\newcommand{\ket}[1]{\lvert #1\rangle}                        % Ket
\newcommand{\bra}[1]{\langle #1 \rvert}                       % Bra
\newcommand{\braket}[2]{\langle #1 \vert #2 \rangle}          % Bra-Ket
\newcommand{\mean}[1]{\langle #1 \rangle}                     % Mean
\newcommand{\ccmean}[3]{\tensor[_{#1}]{\mean{#2}}{_{#3}}}     % Pre- and Postconditioned Mean
\begin{document}
\title{Classical Field Approach to Quantum Weak Measurements}
\author{Justin Dressel}
\affiliation{Center for Emergent Matter Science, RIKEN, Saitama, 351-0198, Japan}
\author{Konstantin Y. Bliokh}
\affiliation{Interdisciplinary Theoretical Science Research Group, RIKEN, Saitama, 351-0198, Japan}
\author{Franco Nori}
\affiliation{Center for Emergent Matter Science, RIKEN, Saitama, 351-0198, Japan}
\affiliation{Physics Department, University of Michigan, Ann Arbor, MI 48109-1040, USA}

\date{\today}

\begin{abstract}
By generalizing the quantum weak measurement protocol to the case of quantum fields, we show that weak measurements probe an effective classical background field that describes the average field configuration in the spacetime region between pre- and post-selection boundary conditions.  The classical field is itself a weak value of the corresponding quantum field operator and satisfies equations of motion that extremize an effective action.  Weak measurements perturb this effective action, producing measurable changes to the classical field dynamics.  As such, weakly measured effects always correspond to an effective classical field.  This general result explains why these effects appear to be robust for pre- and post-selected ensembles, and why they can also be measured using classical field techniques that are \emph{not} weak for individual excitations of the field.
\end{abstract}

%\pacs{03.65.Ta,03.67.-a,02.50.Cw,03.65.Fd}

\maketitle

%\tableofcontents
Quantum weak measurements \cite{Aharonov1988,*Duck1989,*Aharonov2008,*Aharonov2010,Kofman2012,*Shikano2012,Dressel2013d,*Dressel2012e} have received significant media attention in the past five years, primarily in the context of optical implementations.  Unlike traditional projective measurements in quantum theory, which strongly perturb the system being measured, weak measurements gently nudge the system to leave it nearly unperturbed by the measurement process.  The price one pays for making such a gentle measurement is that the detector signal becomes ambiguous, or noisy \cite{Dressel2010,*Dressel2012b}, so many more measurements are needed to overcome the statistical uncertainty.  

In spite of this limitation, however, there is a distinct advantage to performing such a weak measurement over a traditional measurement.  Due to the minimal perturbation, a second measurement can be made after the weak measurement that will probe nearly the same preparation.  Correlating the results of the first weak measurement and the subsequent measurement thus enables access to otherwise inaccessible information.  

As an example, the wave-like coherence of a preparation can be largely preserved and manipulated to engineer ``super-oscillatory'' interference patterns \cite{Calder2005,*Berry2006,*Ferreira2007,*Aharonov2011,*Berry2012,*Berry2013} in a weakly coupled detector signal. Surprisingly, such interference oscillates faster than the largest Fourier component initially present in the detector, so can be used to amplify its sensitivity.  Moreover, the weakness of the measurement can make this amplification resilient to common technical background noise (e.g., electronic $1/f$ noise) \cite{Starling2009,*Feizpour2011,*Jordan2014, *Ferrie2014}.  As such, this technique has been used successfully to resolve Angstrom-scale optical beam displacements \cite{Hosten2008,*Dixon2009,*Turner2011,*Hogan2011,*Pfeifer2011,*Zhou2012,*Gorodetski2012}, and similarly-small frequency shifts \cite{Starling2010}, phase shifts \cite{Starling2010b}, temporal shifts \cite{Brunner2010,*Strubi2013}, velocity shifts \cite{Viza2013}, and even temperature shifts \cite{Egan2012} to extraordinary precision using modest laboratory equipment.

For another example, a weak measurement of the momentum largely preserves the coherence with position, so correlating averaged weak measurements of momentum with subsequent position measurements can directly determine a locally-averaged momentum vector-field \cite{Wiseman2007,*Traversa2013}.  Kocsis \emph{et al.} \cite{Kocsis2011} implemented such a measurement on an optical beam passing through a two-slit interferometer, which correctly produced the local momentum streamlines predicted by Madelung's hydrodynamic approach \cite{Madelung1926,*Madelung1927} and Bohm's causal approach \cite{Bohm1952a,*Bohm1952b} to the quantum theory, as well as those predicted by the relativistic energy-momentum tensor of field theory \cite{Hiley2012} and the Poynting vector-field of classical electromagnetic theory \cite{Bliokh2013a,Bliokh2013}.  

In a similar vein, Lundeen \emph{et al.} \cite{Lundeen2011,Lundeen2012,*Salvail2013,*Malik2014,*Lundeen2014} demonstrated that weakly measuring correlations between conjugate quantities was sufficient information to directly determine the preparation state itself.  Using a similar tactic, Wiseman \emph{et al.} \cite{Wiseman2003,*Mir2007} showed how these correlations could be used to determine the changes made to a preparation by an intermediate perturbation, which has since been used to verify error-disturbance and complementarity inequalities similar to Heisenberg's uncertainty relation \cite{Lund2010,*Rozema2012,*Weston2013,*Baek2013,*Kaneda2013,*Busch2013,*Dressel2014}.

A general criticism of these experimental results is that they can be obtained equally well using classical electromagnetic fields (e.g., \cite{Bliokh2013a,Ritchie1991,*Howell2010}), so the insistence upon using the quantum formalism to understand the effects may seem forced.  Indeed, with the exception of the few notable experiments that exploit multi-particle correlations using entangled photon pairs (e.g, \cite{Dressel2011,*Pryde2005,*Goggin2011}), the effects can be described using a manifestly single-particle theory.  Moreover, many repeated measurements are statistically required to compensate for the added noise of a weak measurement, so the experiments require conditions essentially equivalent to a classical field limit of the underlying quantum theory.  For photons, this limit produces classical electromagnetic theory \cite{Smith2007,*Tamburini2008,*Bialynicki-Birula2013}.

\begin{figure}[t]
  \begin{center}
    \includegraphics[width=\columnwidth]{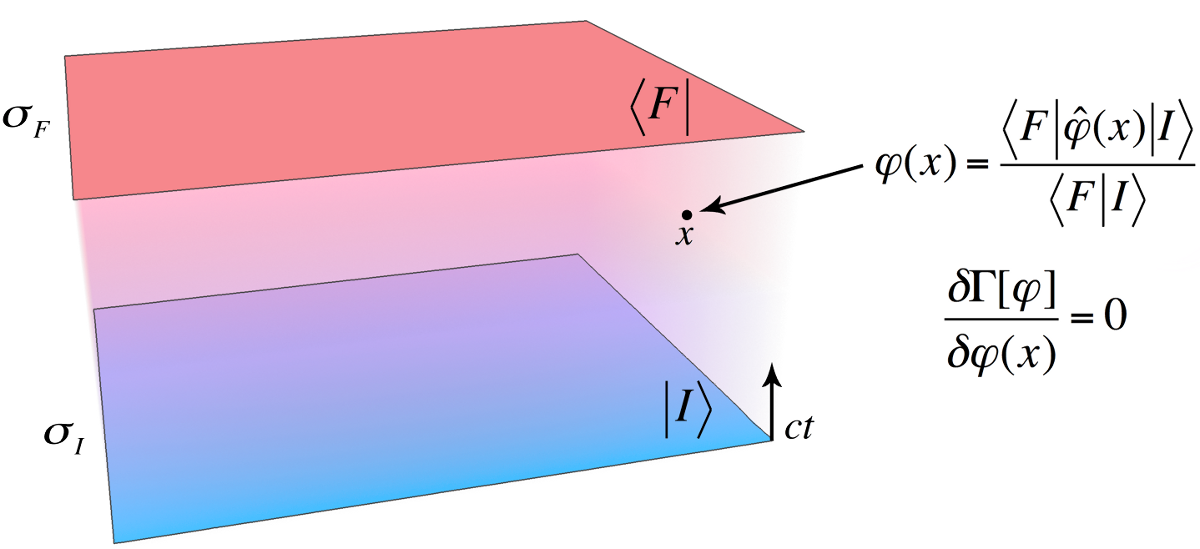}
  \end{center}
  \caption{(color online) Classical background field:  Given two spacetime hypersurfaces $\sigma_I$ and $\sigma_F$, on which field states $\ket{I}$ and $\bra{F}$ are defined as boundary conditions, the classical background field $\varphi(x)$ has the form of a weak value of the quantum field operator $\op{\varphi}(x)$.  It satisfies a classical equation of motion obtained by extremizing an effective action $\Gamma[\varphi]$.}
  \label{fig:background}
\end{figure}

In this Letter we make the connection between weak measurements and classical fields precise.  Specifically, we demonstrate that any weak interaction will probe an effective background field that has the form of the weak value of a local quantum field operator, as illustrated in Fig.~\ref{fig:background}.  The initial and final states for this weak-valued field are defined on spacelike hypersurfaces and provide \emph{boundary conditions}.  Within the bounded spacetime region, the background field deterministically evolves to minimize an effective classical action that satisfies those boundary conditions.  As such, the seemingly ``retrocausal'' character sometimes attributed to weak values (e.g., \cite{Aharonov2010}) originates from precisely the same teleology that underscores the celebrated principle of extremized action.

It follows that averages of weak measurements subject to specific boundary conditions will produce values associated with a corresponding classical background field.  This work complements and explains the observation in \cite{Bliokh2013a,Bliokh2013} that measuring weak values of photon observables will identically recover the observable values of the classical electromagnetic field.  Importantly, this result also implies that the conditions for making a weak measurement may be considerably generalized: any measurement that does not appreciably perturb the classical background field or its boundary conditions will produce the same result as a quantum weak measurement, whether or not the measurement coupling is weak for every field excitation.

\emph{The Quantum Action Principle}.---As a brief review, the essential dynamical principle for quantum fields can be elegantly expressed using Schwinger's variational principle for transition amplitudes \cite{Schwinger1966,*Schwinger1951,*Schwinger1951a,*Schwinger1951b,*Schwinger1953,*Schwinger1953a},
\begin{align}\label{eq:schwinger}
  \delta\mkern-2mu \braket{F}{I} &= \frac{i}{\hbar}\bra{F}\delta\mkern-2mu\op{S}\ket{I}.
\end{align}
Here $\delta$ expresses a variation, $\delta\mkern-2mu\op{S}$ is any Hermitian variation of the quantum action in operator form, and $\ket{I}$ and $\bra{F}$ are specific initial and final field states.  These states are defined on spacelike hypersurfaces $\sigma_{I}$ and $\sigma_{F}$ (i.e., initial and final times) to provide \emph{boundary conditions} for local fields in the interior, as illustrated in Fig.~\ref{fig:background}.  The remaining boundaries for the spacetime volume are assumed to extend to infinity in the space-like directions, where the fields are assumed to vanish.

For collider experiments one typically uses this relation to calculate scattering matrix amplitudes with boundaries that also asymptotically approach infinity in the timelike direction.  These scattering amplitudes are usually expressed in terms of vacuum-to-vacuum amplitudes that are calculated perturbatively from known asymptotically-free solutions.  However, it is worth noting that the dynamical principle of Eq.~\eqref{eq:schwinger} applies generally even outside these scattering conditions.  Indeed, Schwinger \cite{Schwinger1966} demonstated how to derive the operator forms of all conserved quantities, their commutation relations, and the equations of motion for quantum electrodynamics solely from this principle.

\begin{figure}[t]
  \begin{center}
    \includegraphics[width=\columnwidth]{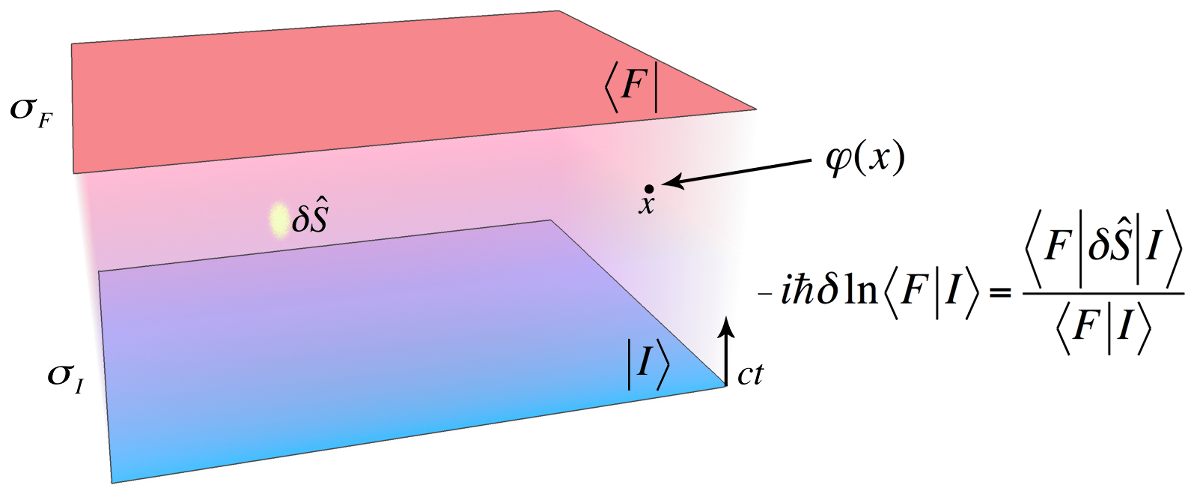}
  \end{center}
  \caption{(color online) Weak measurement:  A weak perturbation of the action $\delta\mkern-2mu\op{S}$ that keeps the boundary states fixed produces a change in the functional $-i\hbar\ln\braket{F}{I}$ equal to the weak value of the perturbation.  The effective action $\Gamma[\varphi]$ is the Legendre transform of this functional, so the classical background field $\varphi(x)$ is correspondingly perturbed.}
  \label{fig:schwinger}
\end{figure}

Under the assumption of local interactions at each spacetime point $(ct, x, y, z)$, Schwinger \cite{Schwinger1966} also showed that variations in the action are additive, so the full variation $\delta\mkern-2mu\op{S}$ connecting the boundaries at $\sigma_{I}$ and $\sigma_{F}$ has the general form,
\begin{align}\label{eq:action}
  \delta\mkern-2mu\op{S} &= \frac{1}{c}\,\delta\!\!\int_{\sigma_I}^{\sigma_F} \!\!\! d^4\!x \, \op{\mathcal{L}}(x),
\end{align}
in terms of a spacetime integral of a local Lagrangian-density $\op{\mathcal{L}}(x)$.  This Lagrangian density must be invariant under the appropriate global and local group symmetries, including the Poincar\'e group that defines spacetime itself.  One can understand Eq.~\eqref{eq:action} as a differential formulation of Feynman's path integral for the amplitude.  Such a variation is illustrated schematically in Fig.~\ref{fig:schwinger}.

The density $\op{\mathcal{L}}(x) = \op{\mathcal{L}}[\op{\varphi}(x),\partial_\mu\op{\varphi}(x)]$ can be further expanded as a functional of local field operators $\bra{x}\op{\varphi}\ket{x'} = \bra{x}\op{\varphi}\ket{x}\,\delta(x - x') = \op{\varphi}(x)\,\delta(x - x')$ and their derivatives, whose specific structure we leave arbitrary here.  Conjugate fields $c\, \op{\pi}^\mu = \partial\op{\mathcal{L}}/\partial(\partial_\mu\op{\varphi})$ can then be introduced to relate the Lagrangian density to a Hamiltonian density $\op{\mathcal{H}}[\op{\varphi},\op{\pi}^\mu]$ using an appropriate Legendre transform.  Integrating this density over a spacelike hypersurface $\sigma$ produces the Hamiltonian operator $\op{H} = \int_\sigma\! d^3\!x\, \op{\mathcal{H}}$ used to generate translations along the timelike coordinate normal to $\sigma$ \cite{Schwinger1966}.

A standard technique to formally compute time-ordered amplitudes of the field operators is to introduce an auxilliary classical source $J(x)$ using a linear variation $\delta\mkern-2mu\op{S} = \int \! \text{d}^4\!x \, \delta\mkern-2mu J(x)\op{\varphi}(x)$ of the Lagrangian density.  After defining the source-dependent amplitude $Z[J] = \braket{F}{I}_J$, it then follows from Eq.~\eqref{eq:schwinger} that
\begin{align}
  g\left(\frac{\hbar}{i}\frac{\delta}{\delta\mkern-2mu J(x)}\right)Z[J] &= \bra{F}\mathcal{T}g(\op{\varphi}(x))\ket{I}_J
\end{align}
for any analytic function $g$ of the field operators, where $\mathcal{T}$ is the time-ordering operation and the right-hand side generally requires regularization.  Similarly, the time-ordered $n$-point correlation functions of the field-operators are generated by the functional $W[J] = -i\hbar \ln Z[J]$ \cite{Schwinger1966} according to
\begin{align}\label{eq:correlations}
  \ccmean{F}{\op{\varphi}_n\cdots\op{\varphi}_1}{I} &= e^{-i W[J]/\hbar}\frac{(-i\hbar)^n\delta^n}{\delta\mkern-2mu J_n \cdots \delta\mkern-2mu J_1} e^{i W[J]/\hbar}.
\end{align}

\emph{Background field}.---The \emph{background field} $\varphi$ associated with a quantum field operator $\op{\varphi}$ is defined as the one-point correlation function from Eq.~\eqref{eq:correlations} \cite{Schwinger1966,Abbott1982,*Vilkovisky1984},
\begin{align}\label{eq:meanfield}
  \varphi(x) &= \ccmean{F}{\op{\varphi}(x)}{I} = \frac{\delta\mkern-2mu W[J]}{\delta\mkern-2mu J(x)} = \frac{\bra{F}\op{\varphi}(x)\ket{I}_J}{\braket{F}{I}_J}.
\end{align}
This background field $\varphi(x)$ is a \emph{classical field} that represents the average field at the point $x$, and is illustrated schematically in Fig.~\ref{fig:background}.  That is, in addition to satisfying the boundary conditions $\ket{I}$ and $\bra{F}$, it satisfies the \emph{classical} equations of motion $\delta\mkern-2mu \Gamma[\varphi]/\delta\mkern-2mu \varphi(x) = -J(x)$, which (in the source-free limit $J\to 0$) extremize an \emph{effective action} $\Gamma[\varphi]$ that is related to the functional $W[J]$ by a Legendre transform $\Gamma[\varphi] = W[J] - \int_{\sigma_I}^{\sigma_F}\!d^4\!x \, J(x) \varphi(x)$ \footnote{Note that we omit technical details regarding the computation of the effective action for gauge fields \cite{Abbott1982}.}.  This effective action can be expanded in powers of $\hbar$ to enumerate the quantum loop contributions to the field dynamics, where the zero-loop contribution can be obtained directly from the quantum action $\op{S}[\op{\varphi}]$ in Eq.~\eqref{eq:action} by replacing the field operators $\op{\varphi}(x)$ with the effective background fields $\varphi(x)$.

In high energy scattering regimes one typically focuses on the vacuum-to-vacuum transition amplitudes between asymptotic infinite times, so the correlation functions of Eq.~\eqref{eq:correlations} reduce to vacuum expectation values and the background field $\varphi(x)$ asymptotically reduces to a free field at the boundaries.  More general plane wave scattering amplitudes can be expressed in terms of these vacuum expectations through a standard reduction procedure.  However, the intermediate interactions can change the structure of the final asymptotic vacuum state from the initial asymptotic vacuum state, leading to distinct initial and final states even for these vacuum-to-vacuum transitions.

\emph{Weak value connection}.---Observe that the last expression for the background field $\varphi(x)$ in Eq.~\eqref{eq:meanfield} has the form of a \emph{weak value} \cite{Aharonov1988} of the local field operator $\op{\varphi}(x)$ with respect to the chosen boundary states.  The classical background field is defined precisely as the weak-valued approximation to a quantum field that applies in the region between the corresponding spacetime hypersurfaces.  This classical background field and its effective action will deterministically describe the average configuration in the interior of the bounded spacetime region.

Importantly, this definition implies that if a local interaction at a point $x$ does not appreciably affect the field dynamics or the boundary conditions, then it will statistically sample the effective classical background field $\varphi(x)$ at that point.  Conversely, since local probes must not appreciably perturb (on average) the dynamics of the background field $\varphi$ or the boundary conditions for the definition in Eq.~\eqref{eq:meanfield} to apply, then these probes must satisfy a weakness criterion to measure $\varphi$ that generalizes the one used by Aharonov \emph{et al.} in \cite{Aharonov1988}.  In particular, the local interaction does not have to be weak for every excited quantum mode of the field; it only has to be weakly perturbing on average with respect to the effective background field to measure the same result.

\emph{Weak measurements}.---To measure the response to an interaction that is weak for every field excitation, as in recent experiments \cite{Kocsis2011,Lundeen2011,Mir2007,Rozema2012,Weston2013}, one can introduce a small variation $\delta\mkern-2mu\op{S} = \frac{1}{c}\int_{\sigma_I}^{\sigma_F}\!d^4\!x \, \delta\mkern-2mu\op{\mathcal{L}}(x)$ in the quantum Lagrangian density itself as illustrated in Fig.~\ref{fig:schwinger} to see how the detection probabilities change.  As in nonrelativistic quantum mechanics, the \emph{normalized} amplitude $a = \braket{F}{I} / \sqrt{\braket{F}{F}\braket{I}{I}}$ will be related to measurable probabilities through a complex square $p = |a|^2$.  Hence, the relative variation in this measurable probability due to the interaction will have the form,
\begin{align}\label{eq:logp}
  \frac{\delta\mkern-2mu p}{p} &= \left[\frac{(\delta\mkern-2mu a^*)}{a^*} + \frac{(\delta\mkern-2mu a)}{a}\right] = -\frac{2}{\hbar}\text{Im}\frac{\bra{F}\delta\mkern-2mu\op{S}\ket{I}}{\braket{F}{I}},
\end{align}
according to Eq.~\eqref{eq:schwinger}, in complete analogy to the situation discussed in \cite{Dressel2013d,Dressel2012d,*Hofmann2011}.  This relation allows one to experimentally measure the imaginary part of the weak value of the perturbation $\delta\mkern-2mu\op{S}$ with respect to the initial and final states of the field by examining logarithmic changes to the detection probability.  

To recover the traditional case of a weak von Neumann measurement used in \cite{Aharonov1988}, consider a variation that is approximately constrained to a spacelike hypersurface $\sigma$ with orthogonal timelike coordinate $ct$.  If the interaction involves two separate degrees of freedom of a local field, $\delta\mkern-2mu\op{\mathcal{L}}(x) = - (\delta\mkern-2mu g)\, \delta(t - t_0)\, \op{\mathcal{H}}_1(x)\otimes\op{H}_2$ with a variable coupling strength $\delta\mkern-2mu g$, and if the initial and final states of the field are product states, then the measurable imaginary joint weak value of Eq.~\eqref{eq:logp} splits into a symmetric sum \cite{Dressel2013d}
\begin{align}\label{eq:wvsplit}
  \frac{\delta\mkern-2mu \ln p}{\delta\mkern-2mu g} &= \frac{2}{\hbar}\left[\text{Re} H_1^w\,\text{Im} H_2^w + \text{Im} H_1^w \text{Re} H_2^w\right],
\end{align}
of both real and imaginary parts of the weak values 
\begin{align}
  H_1^w &= \frac{\bra{F_1}\op{H}_1(\sigma)\ket{I_1}}{\braket{F_1}{I_1}}, & H_2^w &= \frac{\bra{F_2}\op{H}_2\ket{I_2}}{\braket{F_2}{I_2}}.
\end{align}
Here $\op{H}_1(\sigma) = \int_\sigma \! d^3\!x\, \op{\mathcal{H}}_1(x)$ is the effective field Hamiltonian that contributes to an effective interaction Hamiltonian $\op{H}_I = (\delta\mkern-2mu g)\,\delta(t-t_0)\,\op{H}_1(\sigma)\otimes\op{H}_2$ in von Neumann form.  Typically $\op{H}_1(\sigma)$ is a transverse momentum operator that generates spatial translations in the field along a direction in the hypersurface $\sigma$, while $\op{H}_2$ is a spin operator for the field.  The translation operator is constructed from the local conjugate fields $\op{\pi}^\mu(x)$ according to $\op{H}_1(\sigma) = \int_\sigma \! d^3\!x\, q_\mu \op{\pi}^\mu(x)$, where the unit vector $q_\mu$ specifies the translation direction.  All four components of the weak values in Eq.~\eqref{eq:wvsplit} can be determined from averaged measurements that resolve $\partial_g \ln p$ by strategically choosing the boundary conditions to isolate each component up to known scaling factors \cite{Dressel2013d}.

\emph{Classical weak measurements}.---Due to the averaging necessary to resolve the relative probability correction of Eq.~\eqref{eq:logp}, the measured result will match that obtained by introducing the small variation directly to the effective action $\delta\mkern-2mu\op{S}[\op{\varphi}] \to \delta\mkern-2mu\Gamma[\varphi]$ of the classical background field $\varphi(x)$ itself.  To see this, note that the perturbation $\delta\mkern-2mu\op{S}$ affects the generating functional $W[J]$ according to
\begin{align}\label{eq:effactpert}
  \delta\mkern-2mu W[J] &= -i\hbar\delta\mkern-2mu\ln\braket{F}{I}_J = \frac{\bra{F}\delta\mkern-2mu\op{S}\ket{I}_J}{\braket{F}{I}_J},
\end{align}
which is precisely the weak value that appears in Eq.~\eqref{eq:logp}.  According to Eq.~\eqref{eq:meanfield}, this perturbation correspondingly alters the classical background field.  Indeed, the Legendre transform of Eq.~\eqref{eq:effactpert} produces the change in effective action $\delta\mkern-2mu \Gamma[\varphi]$ that alters the equations of motion for $\varphi(x)$.

\emph{Classical electromagnetism}.---As a poignant example, classical electromagnetism can be considered a special case of Eq.~\eqref{eq:meanfield} when the boundaries are \emph{coherent states}, or eigenstates of the positive frequency part of the field operator $\op{F} \propto \op{E} + i c \op{B}$ \cite{Smith2007,Bialynicki-Birula2013}.  Typically, the initial polarization state is assumed to be pure and uncorrelated with the field state, while the final state is left unspecified and thus averaged over all possibilities.  In this special case, Eq.~\eqref{eq:meanfield} produces the classical electromagnetic field $\vec{F} = \mean{\op{F}}$ as an eigenvalue of the field operator with a definite vector orientation of the polarization determined from the initial state.  The effective action $\Gamma[\vec{F}]$ is the classical electromagnetic field action when the loop corrections are neglected; however, it generally contains additional nonlinear corrections when the loops are included \cite{Battesti2013}.  Moreover, the photon number uncertainty in the coherent boundary conditions implies that individual photons may be absorbed by local probes without appreciably perturbing the average background field dynamics, which makes the classical background field description particularly robust in practice.  

Optical experiments that determine the Poynting vector-field by measuring the momentum transfer to small probe particles (e.g., \cite{Oneil2002,Curtis2003,GarcesChavez2003,Bliokh2013a}) are an example of a classical weak measurement.  For each individual photon in the quantum field such a local interaction is not weak: the photon gets absorbed and rescattered.  However, the cross-section of each probe particle is so small that the classical background field is essentially unperturbed by these interactions.  Hence, the averaged interactions measure the local orbital momentum of the classical field \cite{Bliokh2013a,Bliokh2013}.  This is in sharp contrast to the direct technique recently employed by Kocsis \emph{et al.} \cite{Kocsis2011}, who used a local interaction that was weak for each individual photon of the quantum field.  Nevertheless, after averaging these weak interactions over an ensemble of individual photons they obtained the local orbital momentum of the same effective classical field \cite{Bliokh2013a,Bliokh2013}.

Similarly, Lundeen \emph{et al.} \cite{Lundeen2011,Lundeen2014} directly measured the classical background field $\vec{E}$ itself by using a local interaction with a birefringent crystal.  Recall that in an anisotropic medium $\op{D} = \epsilon(\op{E}) = \partial \mathcal{L} / \partial \op{E}$, where $\epsilon$ is the dielectric function and $\mathcal{L}$ is the effective Lagrangian of the medium \cite{Berestetskii1982,Landau1984}.  The relationship between $\op{D}$ and $\op{E}$ determines the birefringence.  For a linear crystal $\op{D} = (\epsilon_0 + \op{\delta})\op{E}$ with a small nondiagonal correction tensor $\op{\delta}$.  Hence, a local birefringence at a point $x'$ originates from a perturbation of the form $\delta\mkern-2mu\op{\mathcal{L}} = \delta(x - x')\op{E}^\dagger(x)\op{\delta}(x)\op{E}(x)$.  For uniaxial birefringence this perturbation is approximately Zeeman-like $\op{\delta}(x) = -i(\delta\mkern-2mu g(x)) \op{S} \partial_{\vec{q}}$, where $-i \partial_{\vec{q}}$ generates translations of the field along a direction with unit vector $\vec{q}$ in the plane transverse to the propagation, $\op{S}$ is a spin operator, and $\delta\mkern-2mu g(x)$ is a local coupling strength \cite{Bliokh2007}.  After expanding the left field operator in the transverse Fourier plane of the hypersurface containing $x$, the perturbation becomes $\delta\mkern-2mu\op{\mathcal{L}} = \delta(x'-x)(\delta\mkern-2mu g) \int\!d^3\!k\, \op{E}_{\vec{k}}\, e^{i\vec{k}\cdot\vec{x}}\,(\vec{q}\cdot\vec{k})\,(\op{S}\otimes\op{E})$.  Hence, choosing $\bra{F}$ to be an eigenstate of the Fourier conjugate field $\op{E}_{\vec{k}}$ (e.g., with a Fourier lens and a pinhole) produces a correction to the effective action according to Eq.~\eqref{eq:effactpert} that is linear in a product $\op{S}\otimes\op{E}$ of the field and spin operators.  As a result, one can measure the classical background field itself up to a constant according to Eq.~\eqref{eq:wvsplit} by strategically choosing the spin boundary conditions.  Measuring the field at each point across the beam profile permits the elimination of a global constant by renormalizing to an effective transverse wave function, as shown by Lundeen \emph{et al.} \cite{Lundeen2011,Lundeen2014}.

\emph{Conclusion}.---We have made precise the connection between locally weak interactions and an effective classical background field.  This background field has the form of a weak value of the corresponding quantum field operator, and evolves deterministically to extremize an effective action while also satisfying the chosen spacetime boundary conditions.  It describes the \emph{average} situation at each local point, but does not describe each field excitation.  

Ideally-weak measurements are noisy, so must be averaged over many realizations.  As such, they probe the properties of this average classical background field, and not the properties of each field excitation.  This observation explains why weak values can also be measured in classical field experiments that don't satisfy the usual criteria of quantum weak measurements.  Each field excitation may be strongly perturbed in these experiments, but as long as the classical background field is negligibly perturbed by the local interaction then the same weakly measured averages will be obtained.  In this precise sense, sufficiently weakly measured quantities can be considered robust properties of a classical background field.

This work was partially supported by the ARO, RIKEN iTHES Project, MURI Center for Dynamic Magneto-Optics, JSPS-RFBR contract no. 12-02-92100, Grant-in-Aid for Scientific Research (S), MEXT Kakenhi on Quantum Cybernetics and the JSPS via its FIRST program.  We thank Edoardo D'Anna and Laura Dalang for their help making figures in this work.

%\bibliographystyle{apsrev4-1long}
%\bibliography{citations.bib}
%merlin.mbs apsrev4-1.bst 2010-07-25 4.21a (PWD, AO, DPC) hacked
%Control: key (0)
%Control: author (8) initials jnrlst
%Control: editor formatted (1) identically to author
%Control: production of article title (-1) disabled
%Control: page (0) single
%Control: year (1) truncated
%Control: production of eprint (0) enabled
%
\end{document}